\documentstyle[11pt,aaspp4]{article}
\input epsf.tex
\begin{document}

\title{Cosmological Constraints from High-Redshift Damped Lyman-Alpha Systems}
\author{Chung--Pei Ma\footnote{
Department of Physics and Astronomy, University of Pennsylvania, 
Philadelphia, PA 19104; cpma@strad.physics.upenn.edu}\,,
Edmund Bertschinger\footnote{Department of Physics, 
Massachusetts Institute of Technology, Cambridge, MA 02139}\,,}
\author{Lars Hernquist\footnote{Presidential Faculty Fellow}
\footnote{Department of Astronomy, 
University of California, Santa Cruz, CA 95064}\,,
David H. Weinberg\footnote{Department of Astronomy, 
Ohio State University, Columbus, OH 43210}\,,
and Neal Katz\footnote{Department of Astronomy, 
University of Washington, Seattle, WA 98195}
}

\def\onu{\Omega_\nu}
\def\qrms{Q_{\rm rms-PS}}
\def\dla{damped Ly$\alpha$\ }
\def\Dla{Damped Ly$\alpha$\ }
\def\go{\mathrel{\raise.3ex\hbox{$>$}\mkern-14mu
             \lower0.6ex\hbox{$\sim$}}}
\def\lo{\mathrel{\raise.3ex\hbox{$<$}\mkern-14mu
             \lower0.6ex\hbox{$\sim$}}}

\begin{abstract}

Any viable cosmological model must produce enough structure at early
epochs to explain the amount of gas associated with high-redshift \dla
systems.  We study the evolution of \dla systems at redshifts $z\ge 2$
in cold dark matter (CDM) and cold+hot dark matter (CDM+HDM) models
using both $N$-body and hydrodynamic simulations.  Our approach
incorporates the effects of gas dynamics, and we find that all earlier
estimates which assumed that all the baryons in dark matter halos
would contribute to \dla absorption have overestimated the column
density distribution $f(N)$ and the fraction of neutral dense gas
$\Omega_g$ in \dla systems.  The differences are driven by ionization
of hydrogen in the outskirts of galactic halos and by gaseous
dissipation near the halo centers, and they tend to exacerbate the
problem of late galaxy formation in CDM+HDM models.  We only include
systems up to the highest observed column density $N\sim 10^{21.8}$
cm$^{-2}$ in the estimation of $\Omega_g$ for a fair comparison with
data.  If the observed $f(N)$ and $\Omega_g$ inferred from a small
number of confirmed and candidate absorbers are robust, the amount of
gas in \dla systems at high redshifts in the $\onu=0.2$ CDM+HDM model
falls well below the observations.
\end{abstract}

\centerline{Submitted to {\it The Astrophysical Journal Letters}}

\section{Introduction}

High-redshift \dla absorbers offer a glimpse of structure in the young
Universe.  These systems dominate the mass density of neutral gas in
the Universe at early times and are believed to be forming galaxies
(Wolfe et al. 1995; Djorgovski et al. 1996).  The fraction $\Omega_g$
of the present-day critical density residing in neutral gas in these
objects increases monotonically with redshift to $z\sim 3.5$
(Lanzetta, Wolfe, \& Turnshek 1995; Wolfe et al.  1995).  At still
higher redshifts, $3.5< z < 4.5$, recent optical spectroscopy of 27
quasars indicates a decline in $\Omega_g$ (Storrie-Lombardi et
al. 1996b).  This behavior is roughly what is expected in a
hierarchical model of galaxy formation: a rise in $\Omega_g$ as
protogalaxies are assembled, then a decline towards low redshifts as
gas is converted into stars.  The precise evolutionary behavior of
$\Omega_g$ depends, however, on the models and can provide strong
constraints on cosmological parameters when compared to the
observations.

Several authors have applied the semi-analytic Press-Schechter (1974)
theory for the mass function of dark matter halos to estimate
$\Omega_g$ in CDM and CDM+HDM models (Mo \& Miralda-Escude 1994;
Kauffmann \& Charlot 1994; Klypin et al.\ 1995).  However, this
analysis depends sensitively on the collapse overdensity parameter
$\delta_c$ and on the assumed range of circular velocities of halos
which are able to host \dla systems.  To reduce the uncertainties
associated with this approach, Ma \& Bertschinger (1994) estimated
$\Omega_g$ and the column density distribution $f(N)$ of \dla
absorbers directly from the dark matter halos in $N$-body simulations.
All of these studies assumed that all the gas in a dark halo becomes
neutral and that the gas mass fraction in each halo is equal to the
global baryon fraction.  This assumption is optimistic, and therefore
conservative from the point of view of ruling out cosmological models.
Nonetheless, all studies found that the then-preferred $\onu=0.3$
version of the CDM+HDM model was ruled out by the observed abundance
of \dla absorbers at $z\sim 3$.  Ma \& Bertschinger (1994) argued that
even an $\onu \sim 0.2$ model would underpredict $\Omega_g$ by a
factor of $\sim 3$.  With hydrodynamic simulations one can calculate
the abundance and properties of \dla systems directly, including the
effects of gas dynamics, radiative cooling, and photoionization.  The
required computations are expensive and complicated, however, and so
far only the standard CDM model has been studied in this way (Katz et
al.\ 1996b, hereafter KWHM).

The goals of this {\it Letter} are to study the consequences of
dissipation and radiative processes for \dla systems in cosmological
simulations and to obtain the strongest possible constraints on
CDM+HDM models from the new data at $3.5 < z < 4.5$
(Storrie-Lombardi et al. 1996abc), which have yet to be considered in
this context.  We use the KWHM hydrodynamic simulation of standard CDM
to compare $f(N)$ and $\Omega_g$ of \dla systems as inferred from the
gas particles directly and from the dark matter alone.  The results
from the CDM model are then used to calibrate our dark-matter-only,
$\onu=0.2$, CDM+HDM simulation.  This approach combines the virtues of
both types of simulations and provides us with the most efficient way
to obtain reliable results that incorporate gas dynamical effects
without performing an expensive hydrodynamic simulation for the
CDM+HDM model.

\vskip -1in
\section{Simulations}

The results presented here for the $\Omega_\nu=0.2$ CDM+HDM model are
based on the particle-particle-particle-mesh (P$^3$M) simulation
reported in Ma \& Bertschinger (1994) and Ma (1995).  The model
assumes $\Omega_{\rm cdm}=0.75$, $\Omega_{\rm baryon}=0.05$, and
$h=0.5$ ($h\equiv H_0/100\,{\rm km}\,{\rm s}^{-1}\,{\rm Mpc}^{-1}$).
The simulation includes a total of 23 million particles ($128^3$ cold
and $10\times 128^3$ hot) in a cubic box of comoving sides
$50\,h^{-1}$ Mpc.  The comoving Plummer force softening is
$25\,h^{-1}$ kpc, and the particle masses are $1.3\times
10^{10}\,h^{-1} M_\odot$ for the cold and $3.3\times 10^8\,h^{-1}
M_\odot$ for the hot particles.  As shown by Ma (1995) and Fig.~1 of
this {\it Letter\,}, increasing the mass resolution of the particles
by a factor of $\sim 10$ has negligible effects on our results.  The
primordial power spectrum has a spectral index of
$n=1$, with density fluctuations drawn from a random Gaussian field.
The normalization corresponds to the 4-year COBE rms quadrupole moment
$\qrms=18\,\mu K$ (Bennett et al. 1996; Gorski et al. 1996), or
$\sigma_8=0.84$ for the rms mass fluctuation in spheres of radius
$8\,h^{-1}$ Mpc.

The hydrodynamic simulation used here was performed with a TreeSPH
code (Hernquist \& Katz 1989; Katz, Weinberg, \& Hernquist 1996a).  We
follow the evolution of structure in a conventional CDM universe with
$\Omega_{\rm cdm}=0.95$, $\Omega_{\rm baryon}=0.05$, and $h=0.5$, with
the Bardeen et al.\ (1986) transfer function for negligible baryon
density.  The simulation volume is a periodic cube of comoving sides
$11.1\,h^{-1}$ Mpc, and it contains $64^3$ gas and $64^3$ CDM
particles.  The comoving gravitational softening length of all
particles is $10\,h^{-1}$ kpc.  The gas and CDM particles have masses
$7\times 10^7\,h^{-1} M_\odot$ and $1.4\times 10^9\,h^{-1} M_\odot$,
respectively.  A uniform ionizing background of
$J_\nu=10^{-22}(\nu_0/\nu) F(z)$ erg s$^{-1}$ cm$^{-2}$ sr$^{-1}$
Hz$^{-1}$ is included, where the redshift dependence $F(z)$ is taken
to be unity for $2<z<3$, $4/(1+z)$ for $3\le z\le 6$, and zero for
$z>6$.  Since the metallicity of intergalactic gas at $z \ga 2$
appears to be 1\% or less of solar (see, e.g., Songaila \& Cowie 1996
and the modeling in Haehnelt, Steinmetz, \& Rauch 1996 and Hellsten et
al.\ 1997), we use a primordial hydrogen-helium composition ($X=0.76$,
$Y=0.24$) to compute cooling rates.  The power spectrum is normalized
to give $\sigma_8=0.7$ today.  This normalization leads to roughly the
observed masses of rich galaxy clusters (White, Efstathiou, \& Frenk
1993), but it is nearly a factor of two below that implied by COBE
(Gorski et al. 1996).  Our goal here is not to investigate the
validity of this model.  Rather, we wish to address the general issue
of the relative distribution of gas and dark matter in galactic halos
and apply the results to the $\onu=0.2$ CDM+HDM model, which comes
closer to matching simultaneously the COBE observations and the
statistics of galaxies and clusters (Ma 1996).

Since we will use the KWHM TreeSPH run to calibrate the expected
effects of gas dynamics in the CDM+HDM N-body run, our analysis
implicitly assumes that the KWHM predictions for damped absorption in
the CDM model are approximately correct, despite the finite resolution
of the simulation.  The physical dimensions of the damped absorbers
are typically $\sim 10-20 h^{-1}\;$kpc (see figure 2 of KWHM),
significantly larger than the gravitational softening at $z \sim 3$
[$\epsilon = 10/(1+z) h^{-1}\;$kpc].  It therefore seems unlikely that
the overall sizes of the systems are much affected by the
gravitational force softening.  The cold gas in the damped absorbers
is almost entirely neutral because of its high density and
self-shielding, so the neutral fraction of this component is
insensitive to resolution.  However, a substantial fraction of the gas
in virialized objects remains hot and produces little absorption, an
effect that is quite important to our results.  The fraction of cooled
gas can be sensitive to mass resolution in some regimes (Weinberg et
al.\ 1997), though a hot gas fraction comparable to that in our halos
is found even in 3-d and 1-d simulations with much higher resolution
(Quinn et al.\ 1996; Thoul \& Weinberg 1996; however, Navarro \&
Steinmetz 1997 find more efficient cooling in their simulations).  The
biggest uncertainty in the KWHM predictions is the inability of such
simulations to model substructure on scales much smaller than the DLA
systems themselves.  We effectively assume that the gas distribution
in damped absorbers is fairly smooth on kpc scales, whereas strong
clumping could systematically shift absorption from low column
densities to high column densities.  Studies of DLA absorbers in
gravitationally lensed quasars may eventually provide empirical
constraints on the coherence of damped systems.

\section{Damped Ly$\alpha$ Systems}

The two characteristics of \dla systems that we choose to examine,
$f(N)$ and $\Omega_g$, are related by 
$\Omega_g=H_0 c^{-1}\mu m_{\rm H}\,\rho_{\rm crit}^{-1}
\int_{N_{\rm min}}^{N_{\rm max}} Nf(N)dN\,$,
where $\rho_{\rm crit}$ is the critical mass density today, $m_{\rm
H}$ is the hydrogen mass, $\mu=1.3$ takes into account that 24\% of
the mass is helium, and $N_{\rm min}=10^{20.3}{\rm cm}^{-2}$.  The
upper limit to the integral could in principle be infinity, but in
practice \dla absorbers have been observed only up to $N_{\rm max}\sim
10^{21.8}{\rm cm}^{-2}$ (Wolfe et al. 1995; Storrie-Lombardi et
al. 1996c).  The absence of observed systems with higher $N$ may
reflect the small-number statistical limitations of current data, or
it may indicate that at higher column densities the gas has been
converted into stars.  The observed $f(N)\Delta N\Delta X$ is given by
the number of absorbers with HI column density in the range
$N\longrightarrow N+\Delta N$ in an absorption distance of $\Delta
X=(1+z)^{1/2}\Delta z$ (for $q_0=0.5$).

The most reliable estimate of $f(N)$ in the CDM simulations comes from
the gas component in the TreeSPH run.  We project the gas particles in
the simulation box onto a two-dimensional grid and compute the neutral
hydrogen column density in each pixel.  Details of the correction for
self-shielding are described in KWHM.  To compare with earlier work,
we compute $f(N)$ from the {\it dark matter} component of the TreeSPH
simulation by projecting each dark matter halo onto a two-dimensional
grid, and converting from dark matter surface density to neutral
hydrogen column density assuming a uniform ratio of neutral gas to
dark matter, $\Omega_b=0.05$, and a hydrogen baryon mass fraction
$X=0.76$.  The dark halos are identified with the
Denmax algorithm (Bertschinger \& Gelb 1991; Frederic 1995),
with no overdensity constraint imposed.

The solid and long-dashed curves in Figure~1 compare $f(N)$ at $z=2$
obtained from the gas and dark matter components in the TreeSPH
simulation.  It shows a steeper slope for $f(N)$ inferred from the
dark matter than from the gas, with the two curves crossing at $N\sim
10^{21.8}$ cm$^{-2}$.  The difference at the high column density end
is most likely due to the cooling of the gas component, which allows
it to contract to a higher density than the dissipationless dark
matter, leading to a higher $f(N)$ in Figure~1.  The difference at the
lower column densities could potentially reflect either different
density profiles of dark matter and {\it total} gas in the outskirts
of galactic halos, or ionization that reduces the {\it neutral} gas
fraction in these regions (or both).  To investigate the two factors
separately, we computed the distribution function for the {\it total}
hydrogen column density in the TreeSPH simulation.  We found an $f(N)$
that closely matches that inferred from the dark matter at column
densities $N \lesssim 10^{22}{\rm cm}^{-2}$, as expected if the dark
matter and the {\it total} gas indeed have similar density profiles
outside the core of each halo.  The gas and dark matter profiles in
Figure 1 of Quinn et al. (1996) also support this result.  Ionization
therefore plays the dominant role in reducing $f(N)$.  Simulations
like these consistently show that cold gas lumps are embedded within
halos of much hotter gas (e.g., Katz, Hernquist, \& Weinberg 1992;
Evrard, Summers, \& Davis 1994), with little gas at intermediate
temperatures.  The cold gas is almost entirely neutral because of its
high density and because of self-shielding.  The large difference
between the dark matter and gas $f(N)$ histograms at $N \sim
10^{20.5}{\rm cm}^{-2}$ in Figure~1 is driven almost entirely by
collisional ionization of the hot gas component.

Two other issues must be addressed before the results of Figure~1 are
applied to $N$-body simulations of the CDM+HDM model.  The first is
the gravitational effect of the dense gas concentration on the dark
matter in the centers of halos.  The baryons, which constitute only
5\% of the mass density in these models, clearly have little dynamical
effect on the global clustering of the halos, but their efficient
cooling in high density regions increases the depth of the
gravitational potential well and can result in a higher concentration
of dark matter near the halo centers in hydrodynamic simulations than
in $N$-body runs.  To quantify this effect, we performed a simulation
identical to the TreeSPH run with the SPH portion turned off (i.e., a
Tree run).  The resulting $f(N)$ from this dark-matter-only run is
shown as a short-dashed curve in Figure~1.  The two histograms for the
dark matter follow each other closely until $N \go 10^{21.8}$
cm$^{-2}$, beyond which the Tree $f(N)$ falls below the TreeSPH run by
more than a factor of two.

The second factor to be considered is the difference in simulation
parameters between the small-box, high-resolution, hydrodynamic CDM
run and the large-box, lower resolution, $N$-body CDM+HDM run.  To
bridge this gap, we compute $f(N)$ (shown as dotted curve in Fig.~1)
from a P$^3$M CDM simulation performed by Gelb \& Bertschinger
(1994a,b), which employed a numerical algorithm and numerical
parameters similar to those of our P$^3$M CDM+HDM run: $33\,h^{-1}$
kpc for the Plummer force softening distance, $50\,h^{-1}$ Mpc box,
and $144^3$ particles.  The agreement among the various curves at
$N\lo 10^{21.6}$ cm$^{-2}$ is remarkable.  At higher column densities,
the P$^3$M curve falls sharply, presumably because of the lower force
resolution.  The difference is, however, only $\sim 50$\% up to the
highest observed column densities ($N\sim 10^{21.8}$ cm$^{-2}$), and
it will be taken into account below.

Having investigated the systematic effects on $f(N)$ of dissipation,
ionization, and simulation parameters in the standard CDM model, we
now turn to the $\onu=0.2$ CDM+HDM model.  Ma \& Bertschinger (1994)
found that $f(N)$ from the large P$^3$M $N$-body simulation of this
model was too steep compared to the data at $z\sim 3$, although they
anticipated that ionization and gaseous dissipation would decrease the
$f(N)$ slope.  Figure~1 now allows us to estimate the correction
needed to ``convert'' dark matter to gas.  We use the ratio of $f(N)$
from the gas component in the TreeSPH run (solid curve in Fig.~1) and
$f(N)$ from the dark matter in the P$^3$M run (dotted curve) as the
correction factor.  We multiply $f(N)$ from the CDM+HDM run by this
factor for each column-density bin to correct for the effects of
ionization and dissipation.  The resulting $f(N)$ is shown in Figure~2
for redshifts $z=2, 3,$ and 4.  Since the observed column density
range extends only to $10^{21.8}$ cm$^{-2}$, the gas correction mainly
acts to reduce $f(N)$ inferred from the dark matter and therefore
worsens the problem of late galaxy formation in CDM+HDM models.  There
is a serious deficit of systems in the $\onu=0.2$ CDM+HDM model in the
highest bin ($10^{21}-10^{21.8}$ cm$^{-2}$) for all 3 redshifts, and a
deficit for all column densities at $z=4$.  It should be noted,
however, that the observed $f(N)$ are based on small numbers of
systems: 12 for the redshift bin $2.5<z<3.5$, and 8 for $z>3.5$
(Storrie-Lombardi et al 1996b).  Moreover, some of the high-redshift
systems await confirmation from high-resolution spectroscopy.  If the
current candidate systems are found to be blending of weaker lines,
the data points will be lowered.  

In the calculation above, we have assumed the same gas to dark matter
conversion factor for $f(N)$ in the CDM and CDM+HDM models.  Although
the HDM component in the latter is less clustered than the CDM
component below the neutrino free-streaming scale, we do not expect
the 20\% HDM present in this model to significantly alter the relation
between gas and dark matter $f(N)$ in Figure~1 for the CDM model.

The neutral gas fraction in \dla systems, $\Omega_g$, is less
constraining than $f(N)$, but it is still an interesting quantity to
measure.  Wolfe et al.  (1995) find $\Omega_g = 0.0051\pm
0.0017\,h_{50}^{-1} $ at $3<z<3.5$, while the combined APM and Wolfe
et al. sample gives $\Omega_g= 0.0030\pm 0.0015\,h_{50}^{-1}$ at
$3<z<3.5$ and $0.0019\pm 0.0008\,h_{50}^{-1}$ at $3.5<z<4.7$
(Storrie-Lombardi et al. 1996b).  Theoretical predictions for
$\Omega_g$ depend sensitively on the upper integration limit $N_{\rm
max}$ if the slope of $f(N)$ is flatter than $-2$ as observed.  KWHM
found the fraction of cold, dense gas in the CDM simulation to be
$\Omega_g\approx 0.0065$, 0.0036, and 0.0017 at $z=2,3$, and 4,
respectively.  However, if we integrate $f(N)$ over the observed
column density range of \dla systems we find much lower $\Omega_g$
values for this model: 0.0011, 0.00096, and 0.00069 at $z=2,3$, and 4.
This drastic reduction arises because the contribution to $\Omega_g$
is dominated by the highest column density systems, which extend only
to $10^{21.8}$ cm$^{-2}$ in the observations but to $10^{24.5}$
cm$^{-2}$ in the hydrodynamic simulation.  To compare fairly with
observations, we choose $10^{21.8}$ cm$^{-2}$ as the upper limit in
$\Omega_g$.  When we use the gas-corrected $f(N)$ to compute
$\Omega_g$ for the $\onu=0.2$ CDM+HDM model, we find
$\Omega_g=0.0004$, 0.00025, and 0.0002 at $z=2,$ 3, and 4,
respectively.  The values at $z=3$ and 4 are nearly an order of
magnitude below the Storrie-Lombardi et al. results.
Gardner et al. (1997b) also finds deficiency in the number of \dla
systems in this model.  

\vskip -.4in
\section{Discussion}
In comparison to the most recent data on \dla systems
(Storrie-Lombardi et al.\ 1996abc), we find the $\onu=0.2$ CDM+HDM
model to underpredict the column density distribution $f(N)$ of \dla
systems at all column densities for $z\sim 4$ and in the column
density range $10^{21}$ cm$^{-2}<N<10^{21.8}$ cm$^{-2}$ for $2<z<4$.
Neither the CDM nor the CDM+HDM model reproduces the observed trend
towards a steeper $f(N)$ at higher redshifts.
We have included the effects of gaseous
dissipation and ionization in our calculations by using the relative
$f(N)$ from gas and dark matter in a CDM TreeSPH simulation to modify
$f(N)$ from the CDM+HDM P$^3$M run.  Without considering cooling and
ionization, one would obtain too steep an $f(N)$ from $N$-body
simulations by assuming that neutral gas traces dark matter.  The dark
matter underpredicts $f(N)$ at higher column densities due to the lack
of dissipation, and it overpredicts $f(N)$ at $N\lesssim 10^{22}$
cm$^{-2}$, primarily because gas ionization on the outskirts of
galactic halos reduces the neutral column densities.

We have also argued that only systems up to the observed limit $N\sim
10^{21.8}$ cm$^{-2}$ should be included in the estimation of
$\Omega_g$ for a fair comparison with data.  All earlier estimates
that assumed all the baryons in dark halos would contribute to \dla
absorption have overestimated $\Omega_g$.  The inclusion of gaseous
processes therefore exacerbates the problem of late galaxy formation
in all CDM+HDM models reported earlier.  However, it should be
remembered that the observed $f(N)$ and $\Omega_g$ are inferred from a
small number of systems, some of which are yet to be confirmed by
high-resolution spectroscopy.

We caution that theoretical predictions of the abundance of galactic
objects at high redshifts are in general very sensitive to the
normalization of the power spectrum.  If the previous 2-year COBE
result $\qrms=20.5\,\mu K$ ($\sim 1.5\sigma$ above the 4-year $18\,\mu
K$) is used to normalize the CDM+HDM model, the predictions in
Figure~2 would be less pessimistic.  However, it is generally the case
that the parameter changes that would alleviate the problems of late
galaxy formation in CDM+HDM (e.g., higher normalization, lower $\onu$,
$n>1$ primeval fluctuation spectrum) would exacerbate its tendency to
produce excessively massive clusters (Ma 1996).

There are certainly limitations to our numerical modeling.  The
hydrodynamic simulation studied in this {\it Letter}, for example,
does not include formation of molecular clouds or stars, which would
reduce the amount of atomic hydrogen available to produce \dla
absorption (Fall \& Pei 1993; Kauffmann \& Charlot 1994).  However,
including star formation and feedback with the simple algorithm
described by Katz et al.\ (1996a) reduces $f(N)$ only at $N\go
10^{22.8}$ cm$^{-2}$ (Gardner et al. 1997a).  The $\Omega_g$ obtained
from systems with $10^{20.3}\,{\rm cm}^{-2} < N < 10^{21.8}$ cm$^{-2}$
therefore may not be much reduced by the process of star formation at
$z\go 2$.  There are additional uncertainties associated with the
finite resolution of the simulations, as discussed in previous
sections.  The TreeSPH simulation, for example, only resolves gas
cooling in halos with $v_c \ga 100 \;{\rm km}\;{\rm s}^{-1}$.  Gardner
et al.\ (1997a) use a combination of numerical and analytic
techniques to estimate that absorption by halos with $v_c \la 100
\;{\rm km}\;{\rm s}^{-1}$ increases the incidence of DLA absorption in
the CDM model by about a factor of two.  Unless the appropriate
resolution corrections are much larger for the CDM+HDM model
considered here, they will not be nearly sufficient to erase the
discrepancy with the Storrie-Lombardi et al. data.

This work was supported in part by the Pittsburgh Supercomputing
Center, the National Center for Supercomputing Applications
(Illinois), the San Diego Supercomputing Center, NASA Theory Grants
NAGW-2422, NAGW-2523, NAG5-2816, NAG5-2882, and NAG5-3111, NASA
HPCC/ESS Grant NAG5-2213, NSF Grant AST 9318185, the
Presidential Faculty Fellows Program, and a Caltech PMA Division
Fellowship.

\clearpage
\begin{figure}
\epsfxsize=5.8truein 
\epsfbox{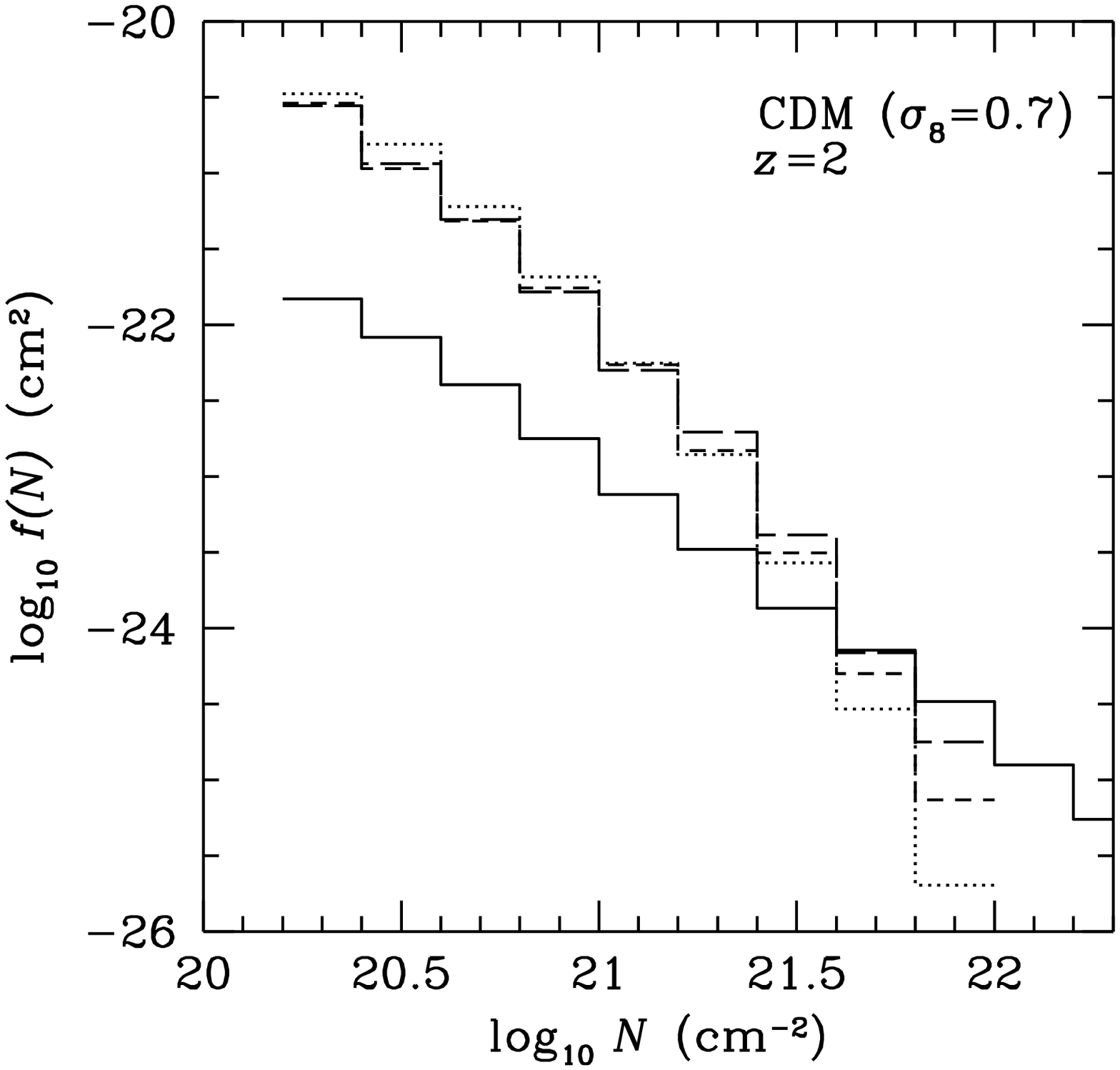}
\caption{
Column density distribution of \dla systems at $z=2$ in the
$\sigma_8=0.7$ CDM model.  The solid histogram is computed from the
{\it gas} component in the TreeSPH simulation (KWHM).  The
long-dashed, short-dashed, and dotted histograms are computed from the
{\it dark matter} component in a TreeSPH, Tree, and P$^3$M simulation,
respectively, under the assumption that gas and dark matter trace
similar density profiles in galactic halos.  The Tree run is identical
to the TreeSPH with the SPH portion turned off; the P$^3$M CDM run has
similar numerical parameters as the P$^3$M CDM+HDM run studied in this
{\it Letter}.  (See text for the exact parameters.)}
\end{figure}
%
\begin{figure}
\epsfxsize=6.8truein 
\epsfbox{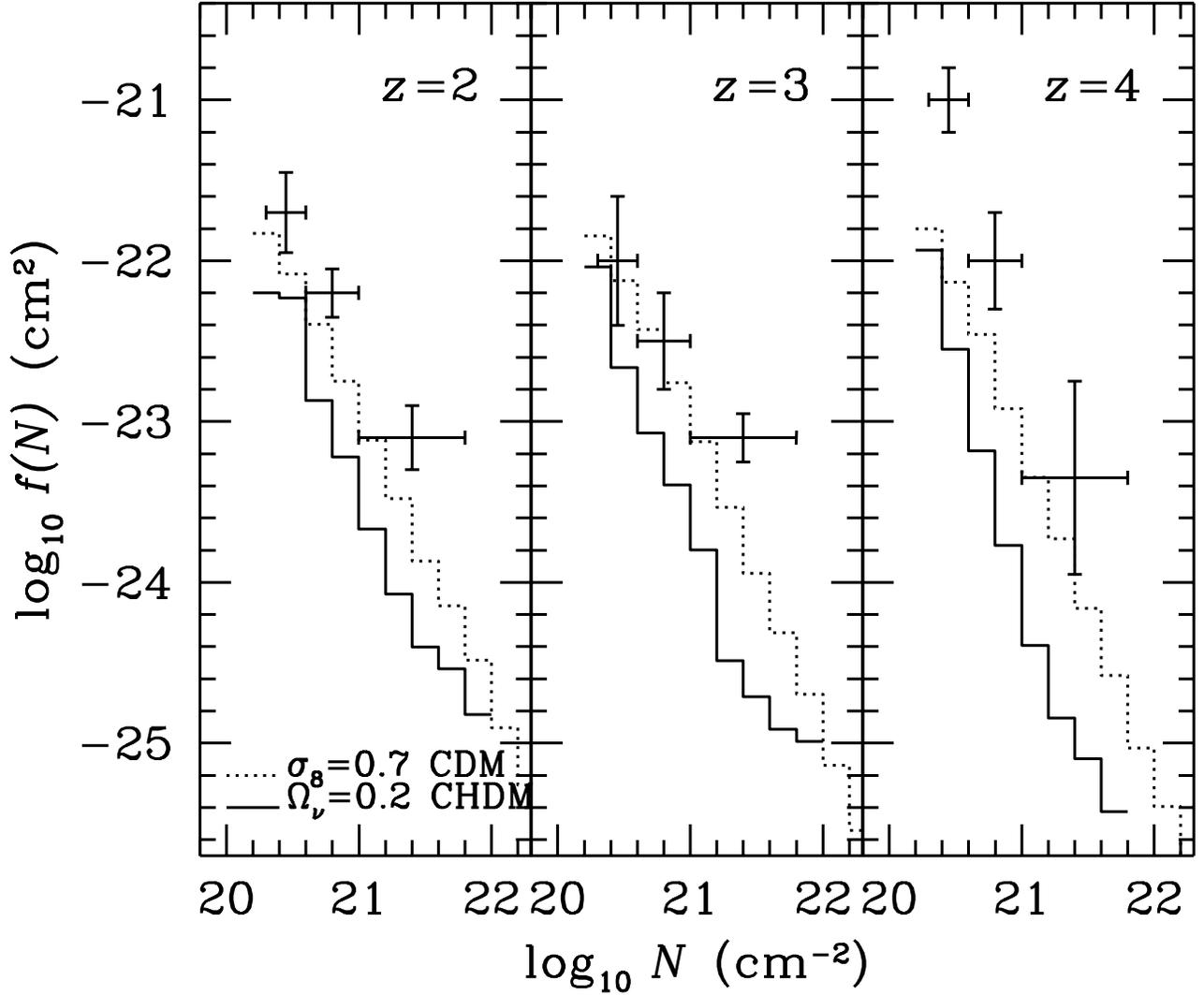}
\caption{
Column density distribution at $z=2$, 3, and 4 from
observations (Storrie-Lombardi et al. 1996a) and models.  The dotted
histogram is computed from the gas component in a TreeSPH simulation
of the low-amplitude CDM model ($\sigma_8=0.7$).  The dashed histogram
is the computed from a P$^3$M simulation of the $\onu=0.2$ CDM+HDM
model normalized to the COBE $\qrms=18\,\mu K$, including the effects
of gas dissipation and ionization.}
\end{figure}

\end{document}